\documentclass[onecolumn]{aastex631}
\usepackage{commath}

\shorttitle{Poynting Flux in the Quiet Sun}
\shortauthors{Tilipman et al.}

\graphicspath{{./}{figures/}}

\received{March 2, 2023}
\revised{June 6, 2023}
\accepted{July 2, 2023}

\submitjournal{ApJ}

\begin{document}

\title{Quantifying Poynting flux in the Quiet Sun Photosphere}

\correspondingauthor{Dennis Tilipman}
\email{dennis.tilipman@colorado.edu}

\author[0000-0001-9361-6629]{Dennis Tilipman}
\affiliation{National Solar Observatory, University of Colorado Boulder, Boulder, CO, USA}
\affiliation{Department of Astrophysical and Planetary Sciences, University of Colorado Boulder, Boulder, CO, USA}

\author[0000-0001-8975-7605]{Maria Kazachenko}
\affiliation{National Solar Observatory, University of Colorado Boulder, Boulder, CO, USA}
\affiliation{Department of Astrophysical and Planetary Sciences, University of Colorado Boulder, Boulder, CO, USA}

\author[0000-0002-5181-7913]{Benoit Tremblay}
\affiliation{High Altitude Observatory, National Center for Atmospheric Research, Boulder, CO, USA}

\author[0000-0002-0189-5550]{Ivan Mili\'c}
\affiliation{Leibniz Institute for Solar Physics (KIS), Freiburg, Germany}
\affiliation{Faculty of Mathematics, University of Belgrade, Belgrade, Serbia}

\author[0000-0001-7764-6895]{Valentin Mart\'\i nez Pillet}
\affiliation{National Solar Observatory, University of Colorado Boulder, Boulder, CO, USA}

\author[0000-0001-5850-3119]{Matthias Rempel}
\affiliation{High Altitude Observatory, National Center for Atmospheric Research, Boulder, CO, USA}

\begin{abstract}

Poynting flux is the flux of magnetic energy, which is responsible for chromospheric and coronal heating in the solar atmosphere. It is defined as a cross product of electric and magnetic fields, and in ideal MHD conditions it can be expressed in terms of magnetic field and plasma velocity. Poynting flux has been computed for active regions and plages, but estimating it in the quiet Sun (QS) remains challenging due to resolution effects and polarimetric noise. However, with upcoming DKIST capabilities, these estimates will become more feasible than ever before. Here, we study QS Poynting flux in Sunrise/IMaX observations and MURaM simulations. We explore two methods for inferring transverse velocities from observations -- FLCT and a neural network based method DeepVel -- and show DeepVel to be the more suitable method in the context of small-scale QS flows. We investigate the effect of azimuthal ambiguity on Poynting flux estimates, and we describe a new method for azimuth disambiguation. Finally, we use two methods for obtaining the electric field. The first method relies on idealized Ohm's law, whereas the second is a state-of-the-art inductive electric field inversion method PDFI\_SS. We compare the resulting Poynting flux values with theoretical estimates for chromospheric and coronal energy losses and find that some of Poynting flux estimates are sufficient to match the losses. Using MURaM simulations, we show that photospheric Poynting fluxes vary significantly with optical depth, and that there is an observational bias that results in underestimated Poynting fluxes due to unaccounted shear term contribution.

\end{abstract}

\keywords{The Sun (1693) --- Solar atmosphere (1477) --- Solar photosphere (1518) --- Solar chromospheric heating (1987) --- Solar physics (1476)}

\section{Introduction}\label{intro}

Quantitative estimates of vertical energy transport in solar photosphere have been limited, yet they are explicitly relevant to many observed phenomena on the Sun, including flux emergence \citep{Cheung2014,Afanasyev2021}, chromospheric and coronal heating \citep{Withbroe1977,Vernazza.1981}, and solar flares and coronal mass ejections \citep{Tziotziou2013,Kazachenko2015,Pomoell2019}. The flux of magnetic energy, i.e. Poynting flux or Poynting vector, defined as the cross product of electric and magnetic fields, has long been considered a primary mechanism for the energy transport from the photosphere to the overlaying atmosphere, but specific magnetically-driven processes and their relative importance have remained somewhat elusive \citep{Steiner2008,Sheylag2012,Liu2012,Welsch2015}. Typically, the flux of magnetic energy is divided into emergence and shear terms. The emergence term arises from advection of magnetic field lines by upward plasma flows, and the shear term (also called wave term) is associated with twisting of the field lines by horizontal flows.

Quantitative investigations of photospheric Poynting flux are a relatively recent development, owing to the fact that the intermediate quantities needed to compute it -- full electric and magnetic field vectors -- are difficult to obtain even from modern state-of-the-art observations. Significant strides have been made in both magnetic field inversions from observed Stokes profiles \citep{delaCruzRodriguez2019,AsensioRamos2019}, and electric field inversions \citep[e.g., ][]{Welsch2015,Fisher2020}. However, most of the quantitative studies of Poynting flux have been constrained to either simulated data \citep{Sheylag2012,Kazachenko2014,Afanasyev2021,Breu2022,Breu2022Corrigendum}, or active regions and plages \citep{Kazachenko2015,Lumme2019}, since in these settings one deals with relatively high polarimetric signal-to-noise ratios (SNR). In particular, \citet{Breu2022} used high-fidelity simulations of the quiet-Sun (QS) photosphere to explain heating in a coronal loop, while \citet{Kazachenko2015} computed Poynting flux from the active region AR 11158 and found it to be sufficient to explain the heating of chromosphere and corona, according to theoretical estimates in \citet{Withbroe1977}. However, an analogous, observation-based study into Poynting flux in QS has not been conducted. \citet{Yeates2014b} and \citet{Welsch2014} put constraints on the coronal energy associated with motions of photospheric footpoints and plage, but they used ideal MHD formulation of Ohm's law and they assumed zero upward advective motion, thereby neglecting the emergence term of Poynting flux. More recently, \citet{Silva22} produced quantitative estimates of QS Poynting flux, but their focus was mostly on the horizontal flux and their method also included several simplifications, such as the idealized Ohm's law and reliance on apparent motions of magnetic field concentrations to obtain velocities transverse to the line-of-sight (i.e. parallel to plane of sky).

The studies of magnetic features in the QS have been few and far between due to both the noisiness of observations and systematic issues. The Sunrise/IMaX balloon-borne probe provides some of the best currently available QS polarimetry \citep{MartinezP2011}, yet even in this data sample, strong linearly polarized light constitutes only about 10\% of the field of view \citep{Kianfar2018}. Furthermore, there is the outstanding problem of magnetic field 180° azimuthal ambiguity, wherein spectropolarimetric inversions of Stokes profiles return two mathematically valid configurations of transverse magnetic field. While many methods of disambiguation have been proposed \citep[for review of some of them and their respective limitations, see e.g. ][]{Pevtsov2021}, none of them have been validated on QS magnetograms. Since full magnetic vector is necessary to compute Poynting flux, the task of disambiguation is necessary.

As a result of these observational and methodological challenges, quantitative investigations into Poynting flux in QS have been limited. At the same time, there will soon be unprecedented observations of QS from the Daniel K. Inouye Solar Telescope (DKIST), which will allow us to improve significantly on spatial resolution, cadence, and/or polarimetric sensitivity \citep{Rimmele2020}. There are also sophisticated methods of computing Poynting flux, which have not yet been tested on QS data. This presents a gap in the current state of this discipline, which this paper seeks to fill. Since QS constitutes the majority of observed photosphere area-wise, it is imperative that we understand the energy flux from it. The goal of this paper is to compute Poynting flux in the QS photosphere. To this end, we use several methods and we apply them to both observational and simulated data, with a focus on the former.

The remainder of the paper is structured as follows: in \S\ref{sec:data} we describe the observational and simulated data we used in this work, in \S\ref{sec:methods} we explain how we obtain Poynting flux and the necessary intermediate quantities -- full velocity, magnetic field, and electric field vectors. In \S\ref{sec:results} we describe Poynting flux estimates from the various employed methods, and in \S\ref{sec:disc} we discuss them. Finally, in \S\ref{sec:conc} we summarize our findings and outline some of the possible future work.

\section{Data}\label{sec:data}

\subsection{Observational Data: IMaX}\label{sec:data_obs}
\begin{figure*}[ht]
    \begin{center}
        \includegraphics[width=0.49\linewidth]{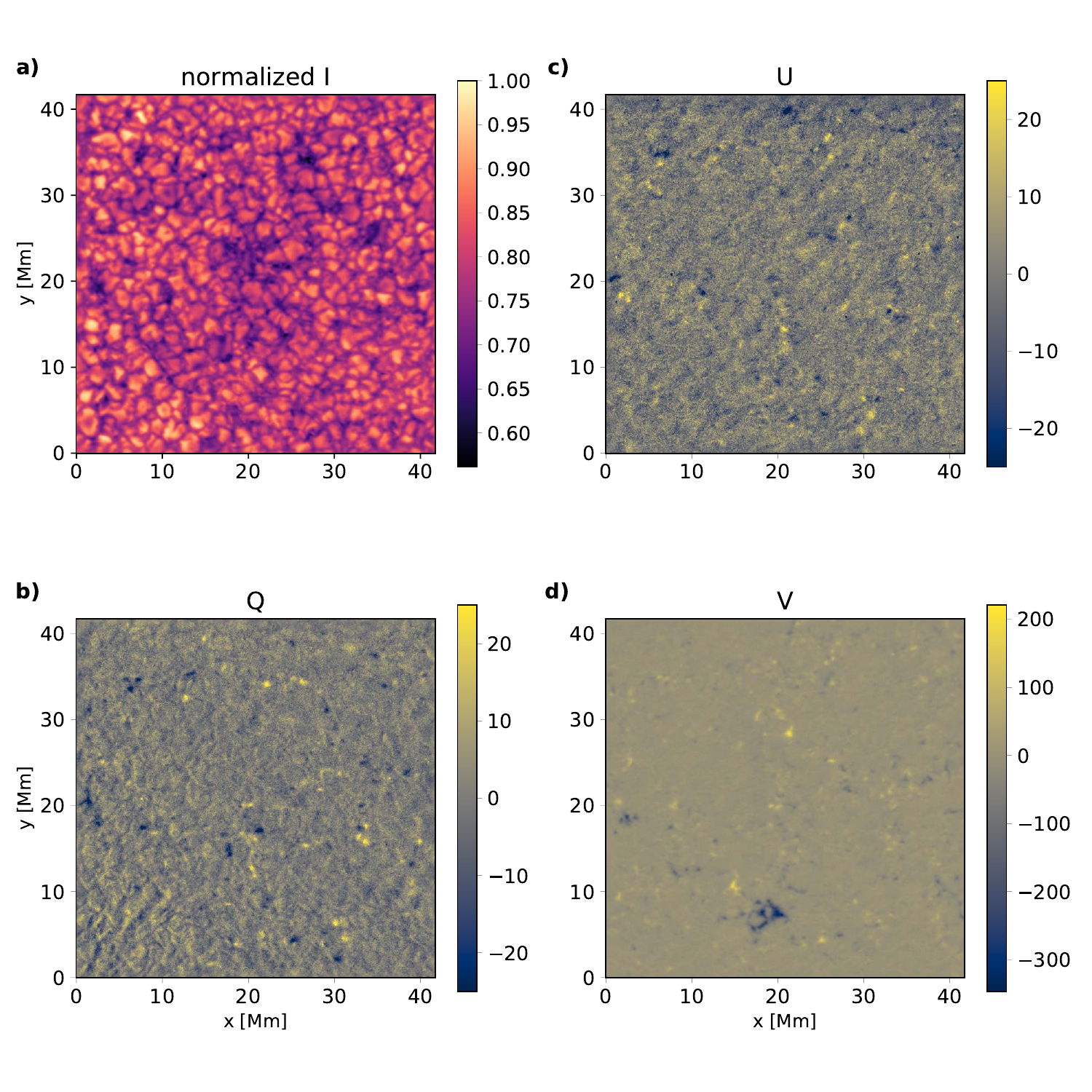}
        \includegraphics[width=0.49\linewidth]{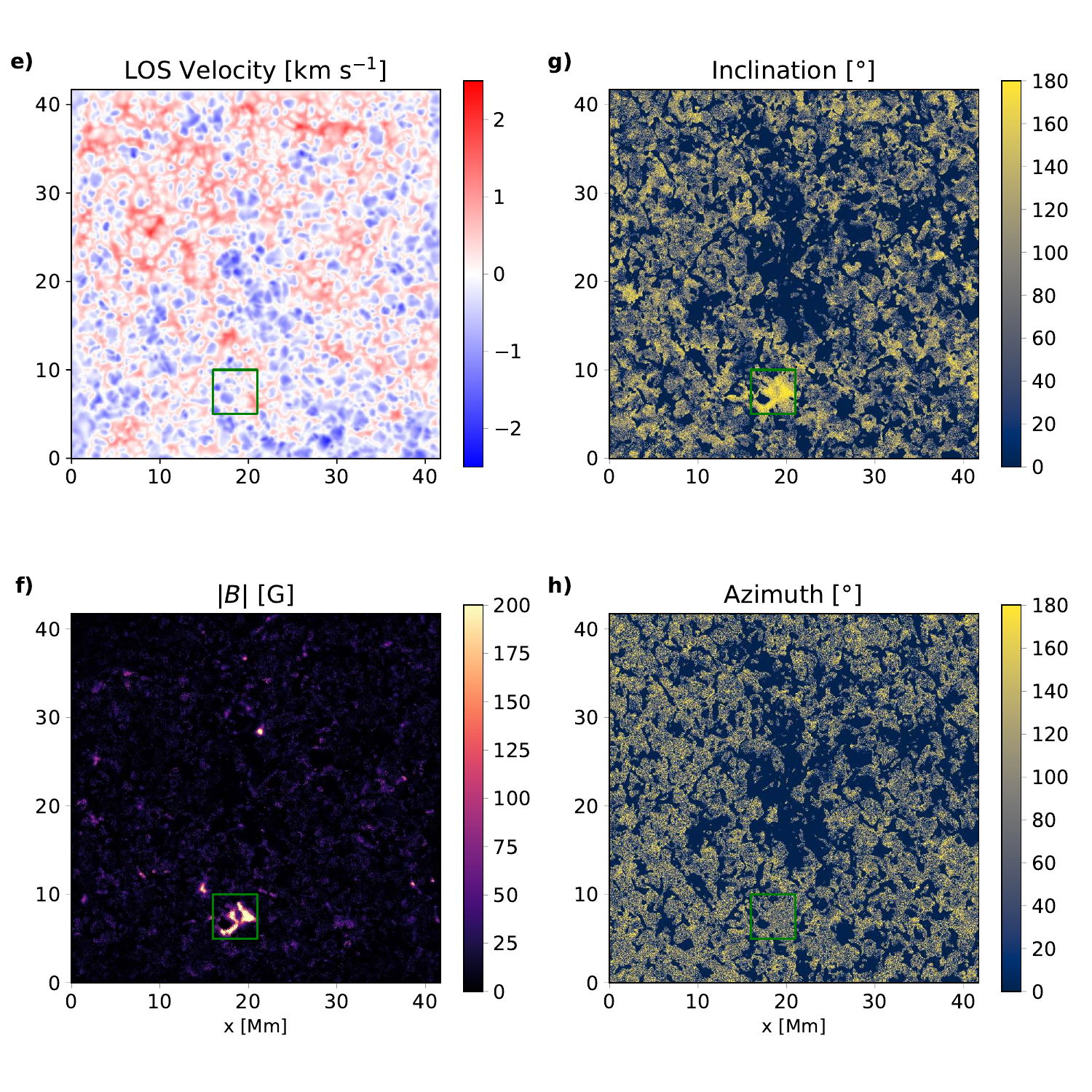}
		\caption{ \small \textit{Left panels, a--d:} IMaX Stokes vectors maps at $t = 1430$ s. \textit{Right panels, e--h:} IMaX Magnetogram and LOS velocity map derived using Milne-Eddington inversions. For LOS velocity, a correction was applied to account for the systematic bias introduced by IMaX Fabry-Perot sensor. The bias scales as a function of distance from FOV center. The green square in the four right panels (e--h) outlines the region of interest (ROI), which contains the strongest magnetic field concentration. A close-up view of ROI is shown in Figure \ref{fig:dismabig}. See \S\ref{sec:Azi} for detailed discussion. The full FOV is $836\times836$ pixels, the ROI is $100\times100$ pixels, and each pixel is 48 km across. } \label{fig:IMaX_lvl1}
    \end{center}
\end{figure*}

We perform our analysis on spectropolarimetric observations from the Imaging Magnetograph eXperiment (IMaX) instrument on board the SUNRISE balloon-borne observatory \citep{MartinezP2011}. We use one continuous IMaX/SUNRISE time series taken on June 9th, 2009 between 01:30:54–02:02:29 UT. This data set covers a $40\times40$ Mm region of QS at the disk center and includes a slowly evolving region of relatively high ($>200$ G, for filling factor unity) magnetic field concentration seen at the bottom of the V Stokes vector map in panels e--h of Fig. \ref{fig:IMaX_lvl1}. The photon SNR of 1000, cadence of 33.25 s, and sampling resolution of 0.0545''/px make the IMaX data set the best available source for the purposes of studying Poynting flux in QS. With this combination of cadence and spatial resolution, a typical flux element moving at a moderate speed of 3 km s$^{-1}$ in the plane of sky \citep[see e.g. ][]{AsensioRamos2017} would traverse two pixels. 

IMaX provides high-quality, diffraction-limited polarimetric observations of QS in the Fe I 5250.2 \AA\ line, which is sensitive to photospheric magnetic fields. The observations include full Stokes vector ($I,Q,U,V$) sampled in five wavelength positions: $\pm40$ and $\pm80$ m\AA\ on either side of the Fe I 5250.2 \AA\ line, and at $+227$ \AA\ in the continuum, with spectral resolution of 65 m\AA\ (85 m\AA\ Gaussian). The IMaX Fabry-Perot sensor introduces a systematic blue shift which grows as a function of distance from the center of field of view (FOV). We apply a correction in the form of distance-to-center-dependent red shift to account for this effect on LOS velocity. The level 0 data had also been corrected to minimize instrumental effects, such as dark and flat-fielding and removal of dust-induced effects, resulting in non-reconstructed (NR) data \citep{MartinezP2011}. The $Q$ and $U$ noise levels in the NR data set were estimated to be $8.3\times10^{-4} I_c$ and $1.1\times10^{-3} I_c$, respectively \citep{Jafarzadeh2014}. In addition, the IMaX point-spread function (PSF) was used to apply a phase diversity reconstruction (PDR) to NR data, thereby increasing spatial resolution to 0.15'' at the expense of increasing $Q$ and $U$ noise levels to $2.6\times10^{-3} I_c$ and $3.6\times10^{-3} I_c$, respectively \citep{Kianfar2018}. We used the NR data, with their lower polarimetric noise, for magnetic field inversions, and the PDR data, with their higher spatial resolution, for velocity inversions.

\subsection{Simulation Data: STAGGER}\label{sec:data_sim1}
STAGGER \citep{Magic2013} is a 3-D radiative magneto-hydrodynamic (MHD) code, which solves for conservation of mass, energy, and momentum equations. Those equations are coupled with radiative transfer equations in local thermodynamic equilibrium (LTE) non-grey atmosphere on a 48-km grid size. The simulation cadence is 60 s. We use continuum intensities and transverse velocities from STAGGER simulations of QS to validate velocities obtained with FLCT and the neural network based method DeepVel, which isdiscussed further in Section \ref{sec:Vinv}.

\subsection{Simulation Data: MURaM}\label{sec:data_sim2}

\begin{figure*}[ht]
    \begin{center}
        \includegraphics[width=0.95\linewidth]{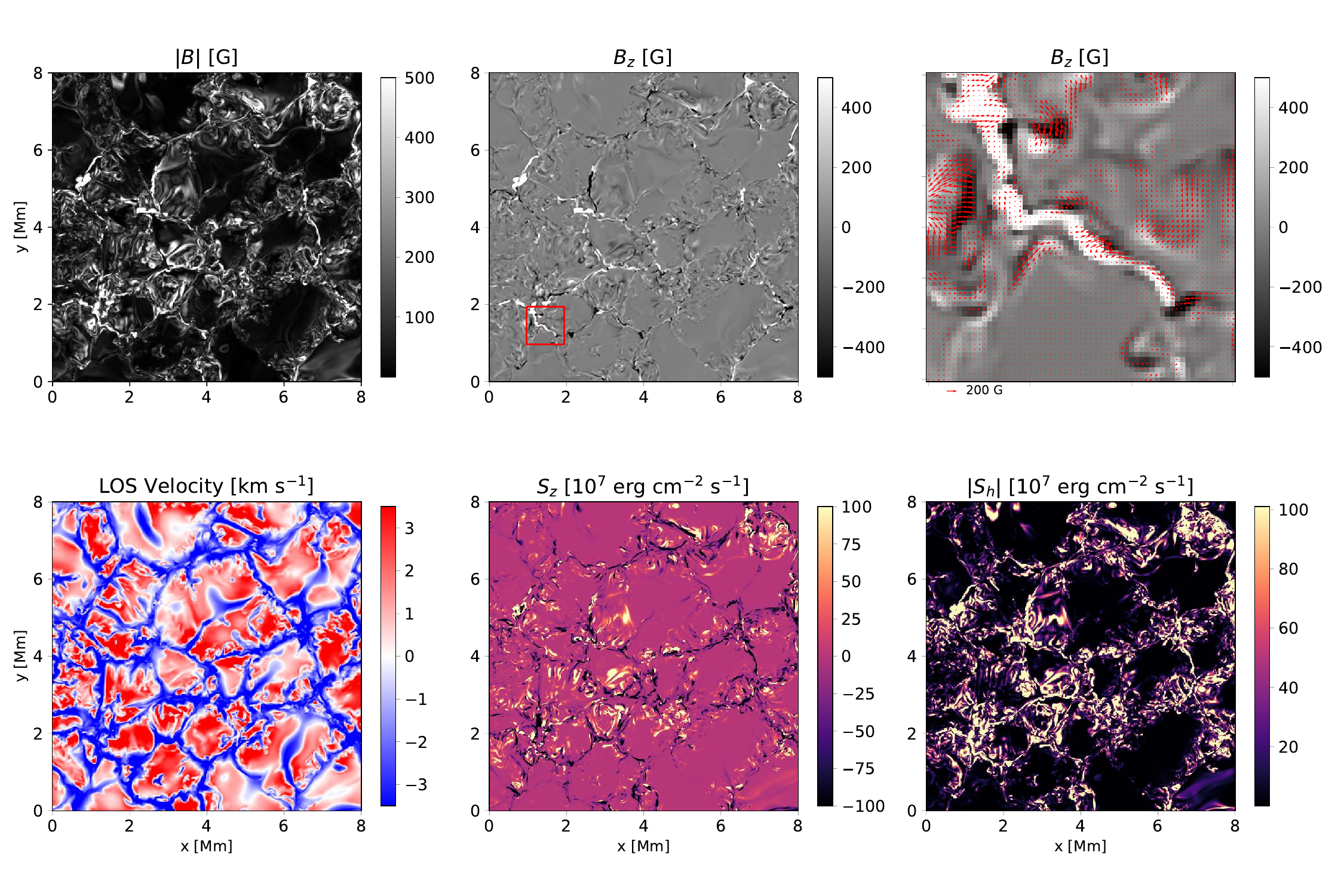}
		\caption{ \small Outputs of a MURaM simulation at the geometrical surface $z=0$, which corresponds to optical depth $\overline{\tau}=1.1$, averaged over FOV. The top right panel is the inset designated by the red square on the top center panel. Arrows represent the orientation of transverse magnetic fields. The vertical and horizontal Poynting fluxes $S_z$ and $S_h$ are computed using the ideal-MHD method (for details, see \S\ref{sec:methods}). }
		\label{fig:muram_all}
    \end{center}
    
\end{figure*}

MURaM \citep{Vogler2005,Rempel:2014:SSD,Rempel2017} is a state the of art radiative MHD code used to model a variety of features in the solar atmosphere and below. The MURaM code solves for mass and energy transfers between subsurface convective zone and the photosphere, chromosphere, and corona. The simulation we analyze here is based on the case `O16bM' from \cite{Rempel:2014:SSD} and was extended in the vertical direction by about 500~km. The simulation solves for all the main MHD quantities (magnetic and velocity vectors, temperature, pressure, heat and energy fluxes) in a domain with the physical extent of $24.576\times 24.576\times 8.192$~Mm$^3$, with an isotropic grid spacing of 16~km, resulting in a $1536\times1536\times512$ grid. It spans optical depths between approximately  $5\times 10^{-8}< \tau <10^9$, i.e. from the convection zone to the upper chromosphere and transition region. The location $\tau=1$ is found about $2$~Mm beneath the top boundary. The relevant simulation quantities from a QS MURaM simulation (LOS velocity, $|B|$, $S_z$, and $S_h$) are shown in Figure \ref{fig:muram_all}.

\section{Methodology} \label{sec:methods}
Recall that Poynting flux is defined as
\begin{equation}\label{eqn:Sdef}
    \textbf{S} = \frac{1}{4\pi} \textbf{E} \times \textbf{B},
\end{equation}
where $\textbf{B}$ and $\textbf{E}$ are magnetic and electric field vectors, respectively. In \S\ref{sec:Binv}, we first describe how we use the polarimetric observations to infer the magnetic field $\textbf{B}$ in the quiet Sun. In~\S\ref{sec:Azi}, we summarize the three methods we use to disambiguate the azimuth of the horizontal magnetic field: ME0 \citep{Leka2009}, random azimuth and Poynting-flux optimization methods. Finally, in~\S\ref{sec:efield}, we  overview the two approaches we use to derive the electric field $\textbf{E}$: the PDFI\_SS electric field inversion method, that solves the Faraday's induction equation
\begin{equation}\label{eqn:Faraday}
    -\nabla \times \textbf{E} = \frac{\partial \textbf{B}}{\partial t},
\end{equation}
and the simplified electric field inversion method that strictly imposes the idealized Ohm's law 
\begin{equation}\label{eqn:ideal}
    \textbf{E} = -\textbf{v}\times \textbf{B},
\end{equation}
where $\textbf{v}$ is the velocity vector. In the simplified formulation of Poynting flux, where idealized Ohm's law is imposed strictly, we can express vertical Poynting flux ($S_z$) as follows:
\begin{equation}\label{eqn:Sz}
    S_z = \frac{1}{4\pi}[v_zB_h^2 - (\textbf{v}_h \cdot \textbf{B}_h) B_z],
\end{equation}
where the $z$ and $h$ subscripts denote vertical and transverse variables, respectively. In this expression, the first term is the emergence term and the second is the wave, or shear, term. In \S\ref{sec:Vinv}, we describe the two transverse velocity reconstruction methods,
DeepVel and FLCT, since transverse velocities are a required intermediate quantity for using either of the electric field inversions. 
For brevity, we refer to the simplified approach as ``ideal-MHD'' method and to the PDFI\_SS method as ``inductive method''. We emphasize, however, that both approaches could enforce the ideal MHD condition, but to different extents: the ``ideal-MHD'' method does so strictly but the ``inductive method'' does not, but could enforce it via ideal non-inductive contribution (see Section 2.4 in \citet{Kazachenko2014})

\subsection{Magnetic Field inversions} \label{sec:Binv}

\begin{figure}[ht]
    \begin{center}
        \includegraphics[width=1\linewidth]{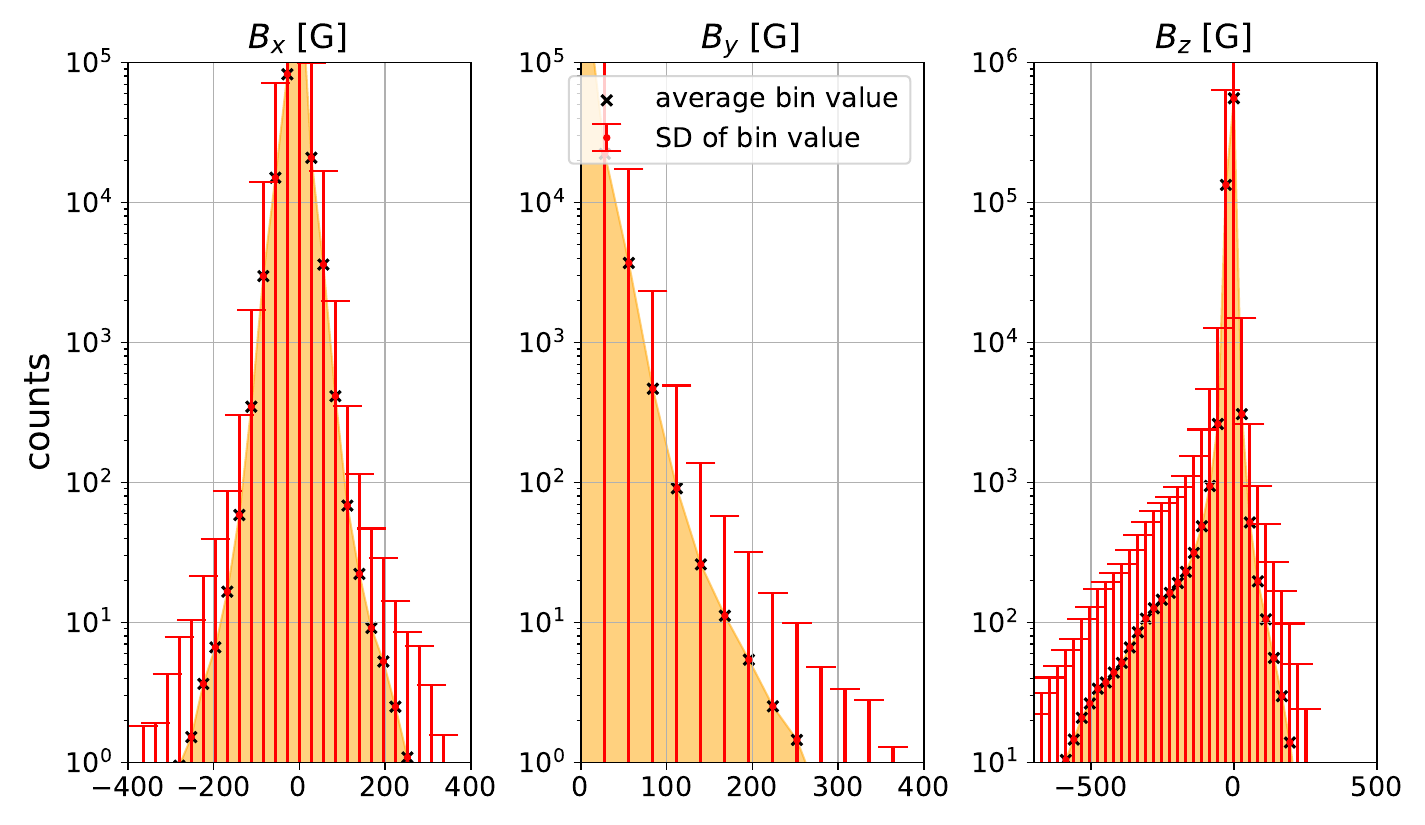}
        \includegraphics[width=1\linewidth]{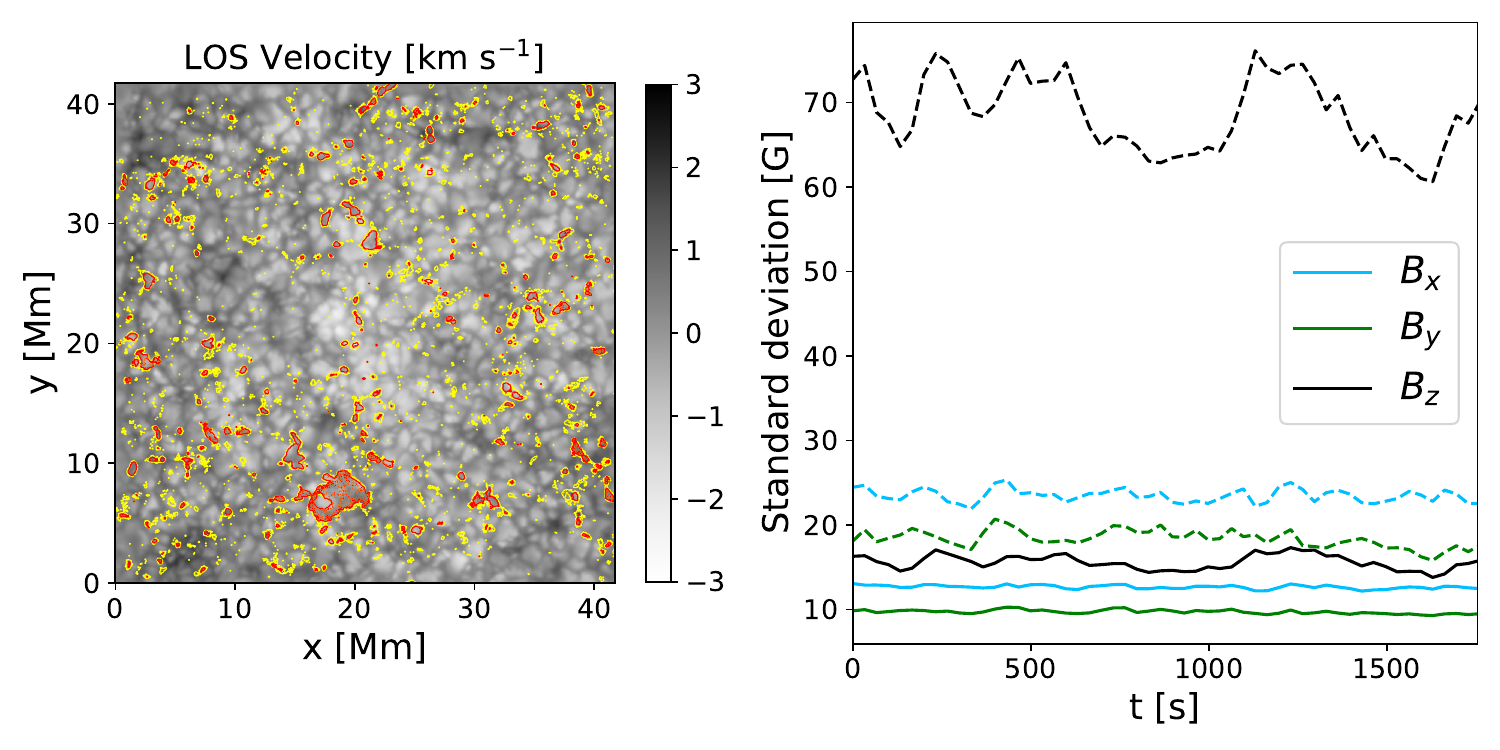}
		\caption{ \small Distributions of \textbf{B}-field components in IMaX magnetograms obtained with Milne-Eddington inversions, before disambiguation at $t=1430$ s. \textit{Top row:} Histograms of \textbf{B}-field components, showing average and standard deviation in each bin taken over all magnetograms during the observation window. The relatively strong negative polarity is evident in the non-symmetric $B_z$ histogram. \textit{Bottom row, left:} LOS velocity map with overlaid contours corresponding to area where polarimetric signal exceeds $3\sigma$ (yellow) and $5\sigma$ (red) at $t=1430$ s. \textit{Bottom row, right:} Standard deviation in \textbf{B}-field components as a function of time taken over the entire FOV (solid lines) and only the pixels with polarimetric signal (at least one of Stokes Q, U, or V vectors) in excess of $3\sigma$ from the continuum (dashed lines). $B_y$ variance is the lowest, since these magnetograms have not been solved for the 180$^{\circ}$ ambiguity. For the same reason, the $B_y$ histogram in the top row does not include negative values. For more discussion on azimuthal ambiguity, see \S\ref{sec:Azi}. }
		\label{fig:Bfield_stats}
    \end{center}
\end{figure}

To obtain the magnetic field configuration and LOS velocity from level-1 IMaX polarimetry, we apply the Milne-Eddington (ME) inversion code pyMilne \citep{delaCruzRodriguez2019} to the NR IMaX data set. We chose this method for its relative computational efficiency -- the assumptions of Milne-Eddington atmosphere simplify the inversion scheme while adequately capturing the physics of the photospheric Fe I 5250.2 \AA\ line formation. b For each IMaX frame, the inversion code uses several seeds (\texttt{nRandom} parameter) to prevent the scheme from converging to local minima, and several Levenberg-Marquardt iterations (\texttt{nIter}) per seed. It should be noted that pyMilne assumes magnetic filling factor of unity, which may introduce bias in transverse magnetic field inversions \citep{Leka2022}.

We show an example of level 1 Stokes data in panels a--d in Figure \ref{fig:IMaX_lvl1} and the corresponding outputs of Milne-Eddington inversions in panels e--h. Clearly visible is the high-V signal region at the bottom of FOV. It corresponds to a strong B-field region that persists and slowly evolves throughout the observation window. We designate it as the region of interest (ROI) and denote it by a green rectangle in the four right panels. The ROI is not associated specifically with either upflows or downflows, as seen from the dopplergram. 

We also show the resulting distributions of LOS and transverse magnetic field components in Figure \ref{fig:Bfield_stats}. The negative polarity in ROI is clearly seen in the skewed shape of $B_z$ histogram. As seen from the bottom-left panel, other regions with strong polarization signal are much smaller in extent. They are also more transient, highlighting the difficulties of QS polarimetric observations. 

As can be seen in Figure \ref{fig:IMaX_lvl1}, the polarimetric signal, particularly in Q and U (panels b and c), is quite weak in our data ($<200$ G in most of the FOV). This is of course to be expected in the quiet Sun regime, where magnetic fields are only strong enough to produce distinct linear polarization features in 3--16 \% of pixels in the FOV of SUNRISE/IMaX \citep{Kianfar2018,Liu2022}. In parts of the analysis that follows, we only consider regions of the FOV with signal strengths above a certain threshold. We chose the threshold of 50 G for masking out pixels with insufficiently strong magnetic fields, for the following reasons: 1) 50 G is approximately equal to $3 \sigma$ in magnetic field strength distribution (bottom right panel of Figure \ref{fig:Bfield_stats}), 2) the subset of pixels in FOV with $B>50$G closely (within 5 G) corresponds to the pixels where at least one of Q, U, or V spectra exhibits strong enough ($>3\sigma$) deviations from continua, 3) this threshold is consistent with the minimum horizontal field strength described in \citet{Kianfar2018}, where the strength of linear polarization features in IMaX magnetograms was found to be in the range 50--500 G.

\subsubsection{Azimuth Disambiguation}\label{sec:Azi}

Azimuthal 180$^\circ$ ambiguity is a well-known problem, wherein spectropolarimetric inversions based on Zeeman effect produce two solutions for $\textbf{B}$-field azimuth, and the two solutions are mathematically equally valid. Several solutions to this problem have been proposed. Those include global optimization mechanisms, such as ME0, where the preferred magnetic field configuration results in a globally minimized magnetic energy \citep{Leka2009}. Other methods select the orientation of magnetic fields that results in the highest $B_z$, if the magnetogram is taken off disk center, or they look for opposite polarities and select an orientation that would close field lines between the polarities \citep{Metcalf.etal2006_rcc}. It should be noted that none of these methods have been rigorously tested in the QS regime, as linear polarization strength is usually too low to adequately employ these methods.

In this work, we attempt to use three (and end up using two) methods to disambiguate azimuths: ME0, randomization, and Poynting flux optimization. We first use ME0, as it is the most physically rigorous of the three methods and it has been extensively used, including, for example, in Hinode and SDO/HMI data processing pipelines \citep{Leka2009Hinode,Hoeksema2014}. ME0, or the minimum energy method, is an optimization algorithm that minimizes the global quantity $\lambda |J_z| + |\nabla \cdot \textbf{B}|$, where $J_z$ is vertical current and $\lambda$ is a modifiable scalar parameter that determines the relative importance of the two terms. As mentioned, ME0 has not been tested on QS data, so, to that end, we tested ME0 on synthetic magnetograms obtained from the 3D MHD STAGGER code (see \S\ref{sec:data_sim1}). Unfortunately, ME0 performed poorly on QS magnetograms produced by STAGGER, likely due to different physical assumptions under which STAGGER and ME0 operate. The issue with ME0 validation warrants a more detailed investigation, but we leave it for future work, as it is not the focus of the present paper.

While we cannot use full ME0 capabilities, the code is capable of performing a potential field acute angle disambiguation. We use this method to disambiguate azimuths in the first frame, and for each subsequent frame, we resolve the ambiguity using acute angle with respect to the previous frame. Another approach is to use the regular ME0 disambiguation while setting the $\lambda$ weighting factor for $|J_z|$ to 0. The minimized quantity is then simply the divergence of magnetic field. Like in the potential field disambiguation, we only apply this method to the first frame and then select the azimuths resulting in an acute angle with respect to the previous frame. In both cases this is done in order to minimize temporal discontinuities. While not strictly physical, this approach has been taken before, e.g. in \citet{Kaithakkal2023}.

In the absence of a validated physical disambiguation method, we asked two questions: how sensitive is Poynting flux to the orientation of transverse magnetic fields (in other words, how much does the "choice" of azimuth affect our computed quantities of Poynting flux), and what is the maximum Poynting flux that can be obtained from any given magnetogram that is yet to be disambiguated? These two questions lead us, respectively, to two other disambiguation methods: azimuth randomization and Poynting flux optimization.

Azimuth randomization can be thought of as an absolutely imperfect disambiguation, wherein we randomly add either 0$^\circ$ or 180$^\circ$ to the azimuth value of each pixel in a magnetogram. The random assignment for each pixel is performed independently of its neighboring pixels or earlier azimuth values in that pixel. Thus, this method yields a disambiguated magnetogram that almost certainly has spatial and temporal discontinuities in transverse field orientations.

The Poynting flux optimization disambiguation method consists of two steps: in the first magnetogram ($t=0$), we disambiguate azimuths in each pixel by selecting the one that results in higher value of $S_z$ as computed using the ideal-MHD method, i.e. using Equation \ref{eqn:ideal}. Then, for each consecutive magnetogram, we select for each pixel the azimuth value that is closer to the azimuth value of that pixel in the previous frame. In contrast with the randomization method, where every pixel is completely independent from both its surrounding pixels and that pixel in adjacent magnetograms in the time series, the Poynting flux optimization method results in some degree of spatial and temporal azimuth continuity, while also providing us a physical ceiling (i.e. upper boundary) for the Poynting flux. We stress, however, that this disambiguation method is only physically meaningful insofar as it provides the ceiling for Poynting flux.

\begin{figure*}[t]
    \begin{center}
        \includegraphics[width = 1\textwidth]{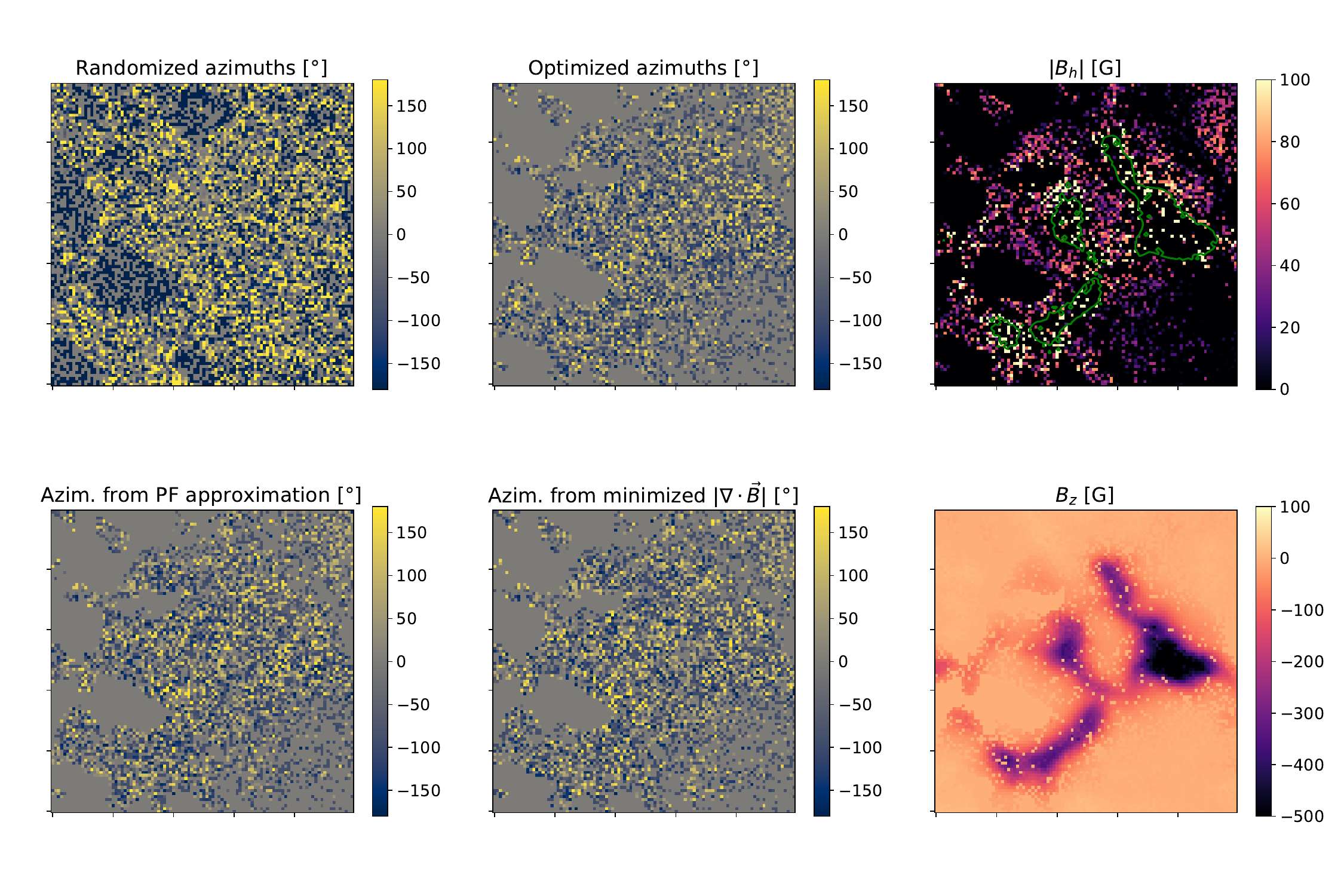}
		\caption{ \small Close-up view of the strong magnetic field region of interest (ROI) at the bottom of FOV (see panels e--h of Figure \ref{fig:IMaX_lvl1}) showing magnetic field azimuthal angles obtained via randomization (top left), optimization (top center), matching to potential field (bottom left), and minimizing B-field divergence (bottom center). The top right panel shows the spatial distribution of horizontal magnetic field in the same region, where green contours correspond to regions with $|\textbf{B}|>200$ G. The bottom right panel shows LOS magnetic field.}
		\label{fig:dismabig}
    \end{center}
    
\end{figure*}

\subsection{Electric Field Inversion Methods}\label{sec:efield}
To find the electric field needed to estimate the Poynting flux (Equation \ref{eqn:Sdef}), we use two approaches. The first ``ideal-MHD'' approach strictly enforces the ideal MHD condition (Equation \ref{eqn:ideal}). We then use Doppler measurements to derive the vertical velocity component and two reconstruction methods, FLCT and DeepVel, to invert the transverse velocity component (see \S\ref{sec:Vinv} below). In the second ``inductive'' approach we use the PDFI\_SS method to derive the electric field directly by inverting Faraday's law without necessarily enforcing the ideal MHD condition (see \S\ref{sec:PDFI} below).

\subsubsection{Transverse Velocity Inversion Methods} \label{sec:Vinv}
As shown in Section \ref{sec:methods}, the full plasma velocity vector (or alternatively, the horizontal electric field) is required to compute Poynting flux. Unlike the LOS velocity, which could be recovered from Doppler data (e.g. \citealt{Welsch2013}), transverse velocities cannot be directly inferred from observables. The two velocity retrieval methods we use in this work are Fourier Local Correlation Tracking \citep[FLCT, ][]{Fisher2008} and a convolutional neural network (CNN) DeepVel \citep{AsensioRamos2017}.

FLCT \citep{Welsch2007} is a plasma flow tracking method that takes two consecutive magnetograms or intensitygrams and, using a finite sliding window, infers the plane-of-sky displacement needed to produce the second map from the first one. It has been used as the flow inversion method for PDFI\_SS \citep{Kazachenko2015,Lumme2019, Afanasyev2021} and for tracking flows in various environments \citep[e.g., ][]{Tremblay2018,Loptien2016}, but it has some constraints. The main constraint of the FLCT approach is that it assumes that any change in continuum or magnetic field intensity is due to advective motion without obeying the induction equation (i.e. FLCT measures an optical flow). Secondly, FLCT has been applied to data with either relatively strong magnetic fields \citep{Welsch2012,Lumme2019}, where tracking is made possible by relatively large S-N ratios, or to low-resolution and large FOV images, where the objective was to track meso- and super-granular motions \citep{Fisher2008}. Neither of these contexts applies to our QS case: magnetic concentrations are, for the most part, transient and limited in spatial extent and strength, making it necessary to rely on continuum images, and the relevant scales of plasma motions are well below even meso-granular scales.

\begin{figure*}[t]
    \begin{center}
        \includegraphics[width = 1\textwidth, trim={0 6.5cm 0 7cm}, clip]{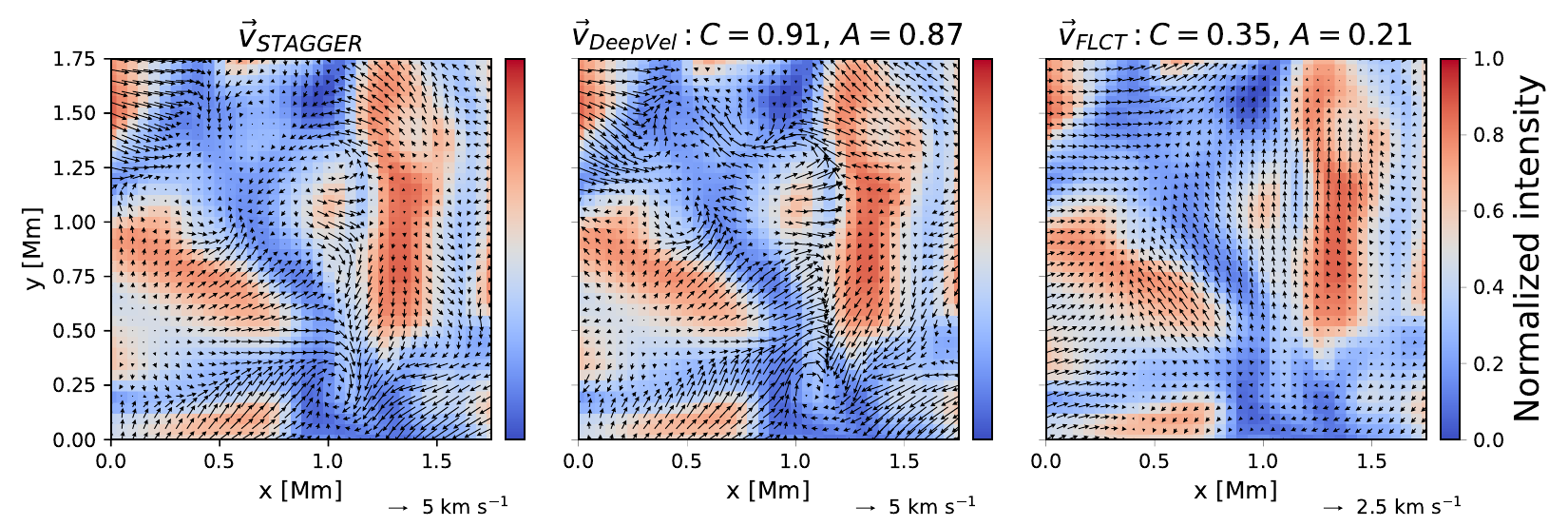}
		\caption{ \small Comparison between the velocity fields computed by the STAGGER simulation (left; reference) and those predicted by DeepVel (center) and FLCT (right) velocity tracking methods. Both FLCT and DeepVel velocity fields were obtained using STAGGER simulated QS intensities, which serve as background for the figure. Note the different scale for FLCT vector arrows.}
		\label{fig:vel_validation}
    \end{center}
    
\end{figure*}

To validate FLCT plasma flow inferences in a setting more closely resembling the IMaX observations, we apply the FLCT to continuum images from STAGGER simulations. We find the correlation between FLCT and reference flows to be low, with the Pearson correlation coefficient of at most $r<0.45$. The correlation is even weaker if the $\sigma$ parameter, which defines the width of sliding Gaussian window, is lower than 10 pixels or higher than 15. In our work, we pick $\sigma=10$, as it produces the strongest correlation. Following analyses in \citet{Schrijver2006} and \citet{Tremblay2021}, we consider three other correlation metrics between reference STAGGER velocities and velocities obtained from inversions: the spatially averaged relative error
$$ E_{\text{rel}}[\textbf{v}_{\text{inv}},\textbf{v}_{\text{ref}}] \equiv \left<\sqrt{\frac{(\textbf{v}_{\text{ref}}-\textbf{v}_{\text{inv}})\cdot(\textbf{v}_{\text{ref}}-\textbf{v}_{\text{inv}})}{\textbf{v}_{\text{ref}}\cdot\textbf{v}_{\text{ref}}}}\right>,$$
the vector correlation coefficient 
$$C[\textbf{v}_{\text{inv}},\textbf{v}_{\text{ref}}] \equiv \frac{\left<\textbf{v}_{\text{inv}}\cdot\textbf{v}_{\text{ref}}\right>}{\sqrt{\left<\textbf{v}_{\text{ref}}\cdot\textbf{v}_{\text{ref}}\right>\cdot\left<\textbf{v}_{\text{inv}}\cdot\textbf{v}_{\text{inv}}\right>}},$$
and the cosine similarity index, which measures the global spatial distribution of velocity vector orientations
$$A[\textbf{v}_{\text{inv}},\textbf{v}_{\text{ref}}] \equiv \left<\frac{\textbf{v}_{\text{inv}}\cdot\textbf{v}_{\text{ref}}}{\norm{\textbf{v}_{\text{inv}}} \norm{\textbf{v}_{\text{ref}}}}\right>,$$ 
where the $<\cdot>$ operation denotes spatial averaging.The $C$ coefficient is defined so that it is $0$ when the velocity vectors are perpendicular everywhere and $1$ when parallel everywhere. Likewise, the $A$ coefficient is $-1$ when the vectors are anti-parallel and $1$ when identical. Thus, the agreement between two vectors is the better the closer both $C$ and $A$ are to unity. For mathematical expressions of these metrics, see equations 3--5 in \citet{Tremblay2021}. For FLCT, the values of these metrics are ($E_{\text{rel}}=1.09,$ $ C=0.35,$ $ A=0.21$). FLCT and reference STAGGER flows are also qualitatively different (see Figure \ref{fig:vel_validation}, left and right panels). STAGGER velocities show a clear pattern of divergence in granules and convergent flows with vortices in intergranular lanes (IGLs), whereas FLCT velocity fields are much more laminar (and smaller in magnitude) on average.

We find that FLCT velocity inferences in QS can be improved by averaging instantaneous velocities over 30-minute time windows \citep{AsensioRamos2017,Tremblay2018}. The correlation coefficient between FLCT and STAGGER velocities then improves to $r=0.75$, but, considering the photospheric timescales are on the order of five minutes, such improvement comes at a cost of losing time-dependent information. We therefore conclude that the FLCT method is an inadequate velocity inversion for our purposes, where instantaneous or near-instantaneous ($<2.5$ minutes) velocities are to have high fidelity.

\begin{figure*}[ht]
    \begin{center}
        \includegraphics[width=0.95\linewidth]{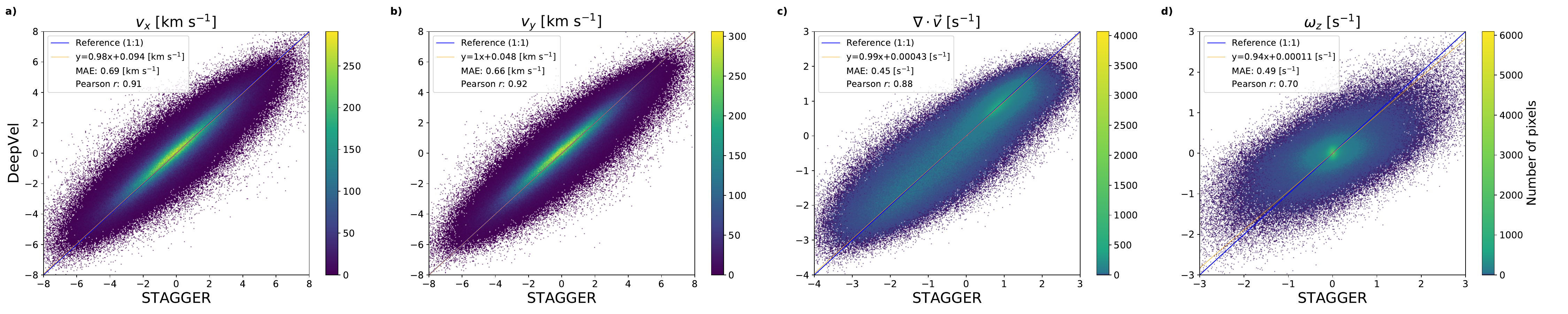}
		\caption{ \small Scatterplots comparing the velocities inferred by the Deepvel neural network to the STAGGER simulation (reference): (a) $v_x$, (b) $v_y$, (c) $\nabla_h\cdot\textbf{v}_h$, and (d) $\omega_z = [\nabla\times\textbf{v}]_z$. Statistical metrics are provided in legends to each panel. MAE stands for mean absolute error.  }
		\label{fig:vel_scatter}
    \end{center}
    
\end{figure*}

In addition to FLCT, we use Deepvel -- a convolutional neural network that has been previously used to infer velocities on granular scales, including in the quiet Sun \citep{AsensioRamos2017,Tremblay2018,Tremblay2020}. DeepVel is trained using simulation data, for which all flow components are known, to map a pair of input images (e.g., continuum intensity images at two timesteps) to the transverse flows at a given optical depth or geometrical height. This approach is known as supervised learning. In other words, the output velocities approximate what the flows in the training simulation would be if we assume that the input data provided to the neural network was generated by the training simulation (i.e., there is a model dependency). For this work, we train DeepVel on a set of STAGGER data frames. To test the trained network, we run it on a STAGGER intensity map that is outside of the training set. The test map has the same properties as those described in Section \ref{sec:data_sim1}. Similarly to the previous works, we find that DeepVel instantaneous velocities are highly ($r=0.91$) correlated with simulated velocities (Figure \ref{fig:vel_validation}).

We find that correlation metrics values are significantly higher for DeepVel ($E_{\text{rel}}=0.74,$ $C=0.91,$ $ A=0.87$) than for FLCT. These are stark improvements over FLCT in terms of the accuracy achieved without losing temporal resolution via time averaging. Even though DeepVel is limited in ways that may have implications for our results \citep[as discussed further in \S\ref{sec:disc} and in e.g. ][]{Tremblay2020}, we choose to apply DeepVel to IMaX intensitygrams to retrieve transverse velocities in our analysis. Hereafter, all retrieved transverse velocities are obtained with DeepVel rather than FLCT. We note, however, that DeepVel is not without its limitations. Even though we get really good agreement ($r\approx0.9$) between simulation and DeepVel velocities and divergences, DeepVel is not as reliable at reproducing vorticities (see Figure \ref{fig:vel_scatter}, panel d).

\subsubsection{PDFI Electric Field Inversion Method} \label{sec:PDFI}
To find Poynting flux without assuming ideal MHD conditions, we use the PDFI\_SS method \citep{Fisher2020}. Briefly, the magnetic field in the PDFI\_SS method is expressed as a sum of poloidal and toroidal components. This decomposition allows to derive the {\it inductive} component of the electric field from observed quantities by uncurling the Faraday law (Equation \ref{eqn:Faraday}). The gradient of a scalar part of the electric field that appears due to uncurling of the Faraday's law is called ``non-inductive'' and could be computed from additional constraints, including ideal MHD constraint $\textbf{E}\cdot\textbf{B}=0$ \citep{Kazachenko2014}. 

PDFI\_SS has been used to describe the evolution of Poynting flux and magnetic helicity in multiple works, but notably, these were all concerned with either observed or simulated active regions \citep[e.g. ]{Kazachenko2015,Lumme2019} or regions of flux emergence \citep[e.g. ]{Afanasyev2021}. To our knowledge, PDFI\_SS has not been applied to QS magnetic fields. Apart from the general challenge of studying QS magnetism, the reliance of PDFI\_SS on the $\displaystyle \frac{\partial \textbf{B}}{\partial t}$ term in Faraday's law makes it especially susceptible to noise. To mitigate the influence of noise, we set the \texttt{bmin} parameter, which masks pixels with lower magnetic field strength \citep[see Section 10.2 in ][]{Fisher2020}, to 50 G -- the same threshold we choose in \S\ref{sec:Binv}. PDFI\_SS also requires high cadence observations, so as to not miss the transient magnetic concentrations that are ubiquitous in QS \citep{Gosic2018}.

\section{Results} \label{sec:results}

\begin{figure*}[ht]
    \begin{center}
        \includegraphics[width=0.49\linewidth]{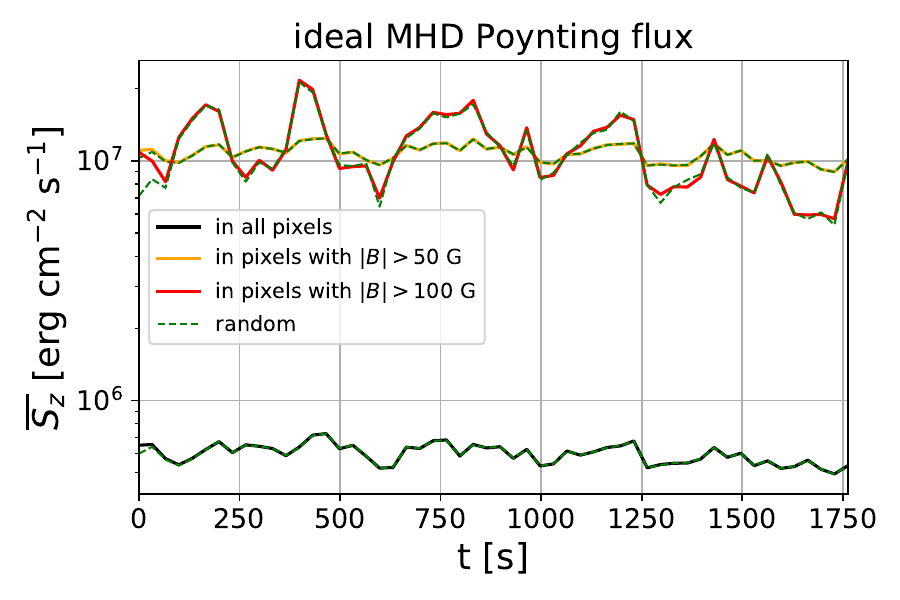}
        \includegraphics[width=0.49\linewidth]{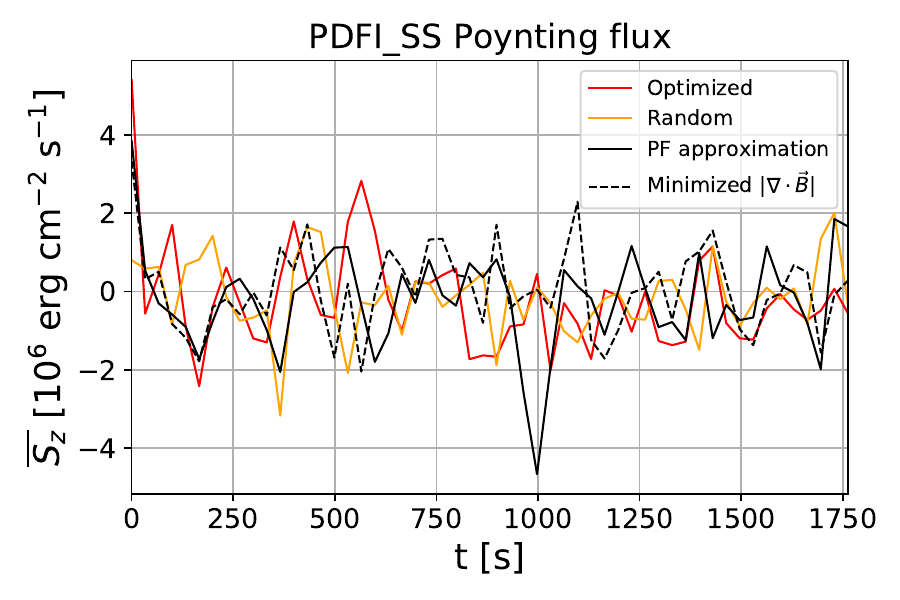}
		\caption{ \small \textit{Left panel:} Temporal evolution of the average Poynting flux, computed using the ideal-MHD assumption and DeepVel velocities. The dashed green lines represent Poynting flux computed in the same sets of pixels as the solid lines, but using random azimuths instead of azimuths obtained via the Poynting flux optimization procedure. \textit{Right panel:} Poynting flux computed via PDFI\_SS and DeepVel velocities and averaged over pixels above the threshold B-field value (see \S\ref{sec:PDFI}), using random azimuths (orange line), optimized azimuths (red line), and azimuths obtained from potential field acute angle method (solid black line) and from imposing $|\nabla \cdot \textbf{B}=0|$ (dashed black line). Note the different y-axis limits. }
		\label{fig:Sz_estimates}
    \end{center}
    
\end{figure*}

We compute Poynting fluxes using two approaches: 
from velocity fields together with the ideal MHD assumption, and the PDFI\_SS electric fields, where time derivatives of magnetic fields are used as a source term. Within each approach, we use randomly disambiguated azimuths and azimuths obtained via the optimization procedure (see Section \ref{sec:Azi}). We show temporal evolution of Poynting fluxes in both settings in Figure \ref{fig:Sz_estimates}.

\begin{figure}[t]
    \begin{center}
        \includegraphics[width=0.8\linewidth]{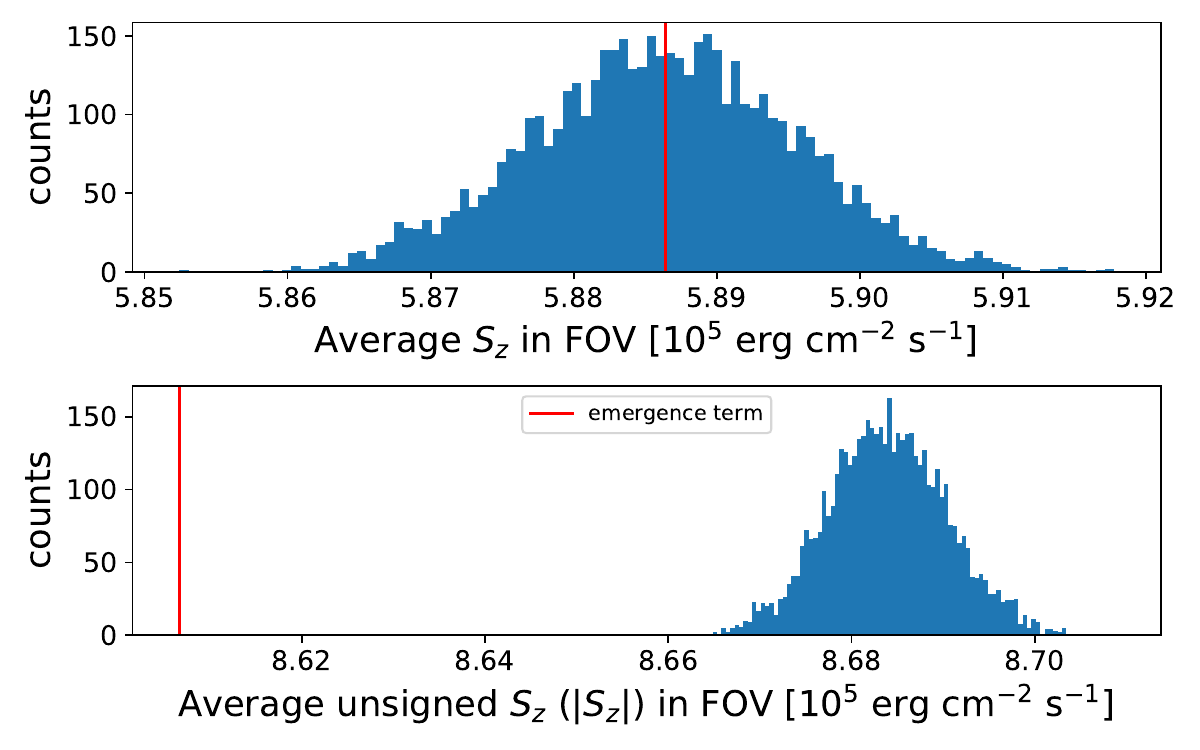}
		\caption{ \small Histograms of the spatially averaged Poynting flux within the FOV at $t=1430$s. The computations were performed using the ideal-MHD method with 5000 realizations of randomized azimuth disambiguation. The vertical red line corresponds to the value of the emergence term of $S_z$ (see \S\ref{sec:results}). }
		\label{fig:randomSzs}
    \end{center}
    
\end{figure}

As discussed in \S\ref{sec:Azi}, azimuthal orientation of vector magnetic fields can affect Poynting flux magnitudes. Since one of our two principal methods of azimuthal disambiguation relies on randomizing azimuths on a pixel-by-pixel basis, we investigate the resulting uncertainty in Poynting fluxes in ideal-MHD setting by repeating the randomization for each magnetogram 5000 times. We find average Poynting flux estimates in each frame to be highly robust to different realizations of azimuth randomization, with both signed (net) and unsigned (absolute values) fluxes tightly clustered (see Figure \ref{fig:randomSzs}). We also find that in the ideal-MHD setting, the emergence term $\displaystyle v_z B_h^2$ dominates both signed and unsigned fluxes over the shear, or wave, term $ ({\textbf{v}}_h  \cdot {\textbf{B}}_h) B_z$ (see Equation \ref{eqn:Sz}), which accounts for less than 1\% of the total vertical Poynting flux.

The left panel of Figure \ref{fig:Sz_estimates} shows the Poynting flux evolution for the ideal-MHD  case. Different plot colors correspond to spatially averaged Poynting flux ($\overline{S_z}$) in all pixels as well as only in pixels where the magnetic field strength ($|\textbf{B}|$) exceeds $50$ G and $100$ G thresholds. The first thing to note here is that the choice of azimuth disambiguation method plays a negligible role on $S_z$ values. The largest difference is in the first frame, where, in the optimization procedure, we explicitly optimize for largest $S_z$ value. In just over one minute this difference disappears, and $\overline{S_z}$ values stabilize within the range $6.0\pm 0.56 \times 10^5$ erg cm$^{-2}$ s$^{-1}$ and $1.1\pm0.087 \times 10^7$ erg cm$^{-2}$ s$^{-1}$  for all pixels and pixels with high $B$-fields, respectively. This is consistent with our analysis of the randomization procedure shown in Figure \ref{fig:dismabig}, where we find very little variation in FOV-integrated $S_z$ across different azimuth realizations. 

We also observe that selecting pixels with relatively strong $|\textbf{B}|$ increases the average $S_z$ by an order of magnitude, but there isn't much variation between 50 G and 100 G thresholds (or even 150 G threshold and above, which aren't shown here). Increasing the threshold to 100 G, however, reveals quasi-periodic oscillations that could conceivably be linked to 5-minute photospheric oscillations \citep{Leighton1962,Ulrich1970}.

Figure \ref{fig:Sz_estimates} (right panel) shows Poynting fluxes derived from the PDFI\_SS method. Recall that, since we set the \texttt{bmin} parameter to 50 G, all pixels with magnetic fields below that threshold are set to zero and are not considered in the following analysis. We find that these estimates are very different from the ideal-MHD estimates shown on the left panel. Firstly, the optimized (randomized) $\bar{S_z}$ is $-2.1\pm 13 \times 10^5$ ($-1.3\pm 9.4 \times 10^5$) erg cm$^{-2}$ s$^{-1}$ -- significantly lower, even in terms of absolute values, than $\bar{S_z}$ in the ideal-MHD case with pixels with weak magnetic fields counted. Secondly, in both cases (randomized and optimized azimuths), $S_z$ oscillates around zero and is sometimes well below it, meaning that magnetic energy is transported downwards instead of upwards. Thirdly and less surprisingly, $S_z$ values obtained from the randomization and $S_z$ optimization disambiguation methods are much more different from one another than in the ideal-MHD case. This is due to the fact that PDFI\_SS uses spatial and temporal derivatives of the $\textbf{B}$-fields to compute $S_z$, and both are affected in the randomization procedure which produces highly discontinuous magnetic field configurations, more so than in the case of optimized azimuths. However, the two other azimuthal ambiguity resolutions -- from potential field and from $|\nabla \cdot \textbf{B}| = 0$ -- also produce Poynting fluxes that oscillate frequently around zero and almost never exceed $2 \times 10^6$ erg cm$^{-2}$ s$^{-1}$. This most likely indicates that significant spatial and temporal discontinuities are present in IMaX QS magnetograms regardless of the azimuthal disambiguation method, as can be seen in Figure \ref{fig:dismabig}. 

To evaluate how Poynting flux and its components vary in height, we use the outputs of MURaM simulations, since IMaX data set only includes data from one optical surface. In Figure \ref{fig:muram_tauS}, we compare Poynting fluxes derived directly from MURaM simulations. We find that MURaM averaged vertical Poynting flux reverses sign very close to the $\tau=1$ surface, and that it is exceeded by $|S_h|$ from the convection zone until well above $\tau=0.1$. We find that at $\overline{\tau}=1$, $\overline{S_z} = 4.38 \times10^6$ erg cm$^{-2}$ s$^{-1}$, and it rises to $2.28 \times10^7$ erg cm$^{-2}$ s$^{-1}$ at $\overline{\tau}=0.1$.

\section{Discussion} \label{sec:disc}

\begin{figure}[ht]
    \begin{center}
        \includegraphics[width=0.9\linewidth]{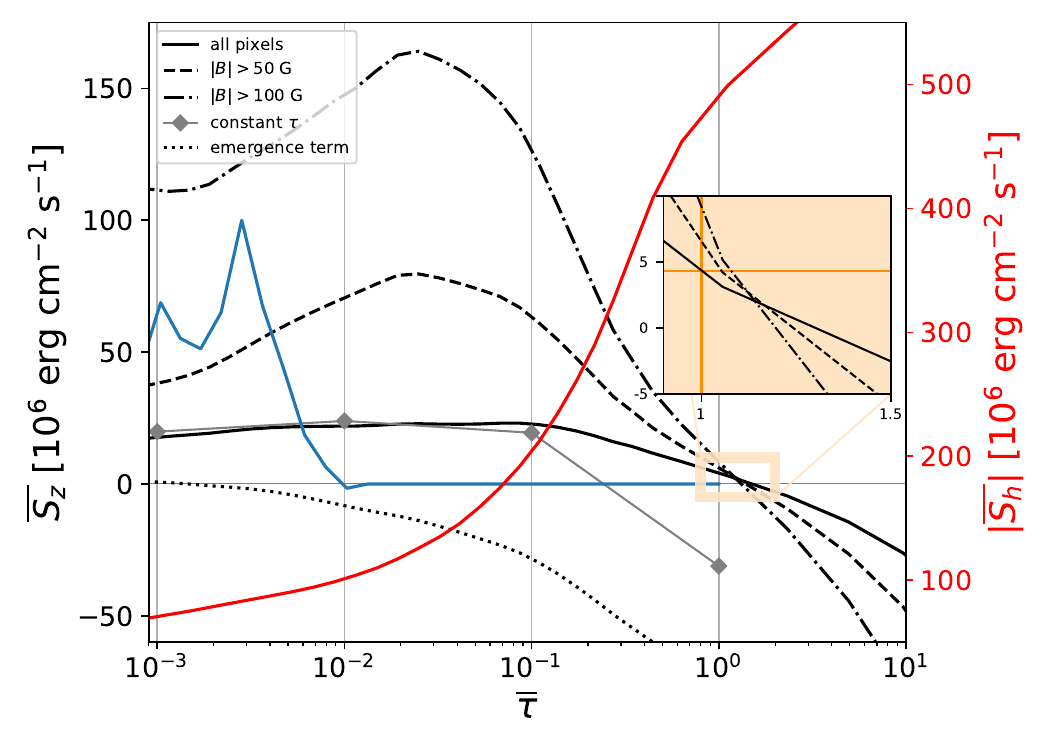}
		\caption{ \small Average Poynting flux in MURaM simulations as a function of optical depth. Red curve shows transverse Poynting flux. The x-axis corresponds to a vertical range of $\sim 0.35$ Mm. The black curves (grey diamonds) correspond to vertical Poynting flux computed on geometrical (optical) surfaces and spatially averaged over subsets of pixels with varying magnetic field strengths. The dotted curve represents the emergence term ($\displaystyle v_zB_h^2$) of the vertical Poynting flux averaged over all pixels on geometrical surfaces. The blue curve represents the response function of the Fe I 5250.2 \AA\ line in IGL in arbitrary units. Its peak is at $\overline{\tau}=3\times10^{-3}$, which is slightly below 400 km. The horizontal and vertical orange lines in the inset represent $S_{z,thr}$ from \citet{Withbroe1977} (see \S\ref{sec:disc}) and the $\overline{\tau}=1$ surface, respectively.}
		\label{fig:muram_tauS}
    \end{center}
    
\end{figure}

\begin{figure}[ht]
    \begin{center}
        \includegraphics[width=1\linewidth]{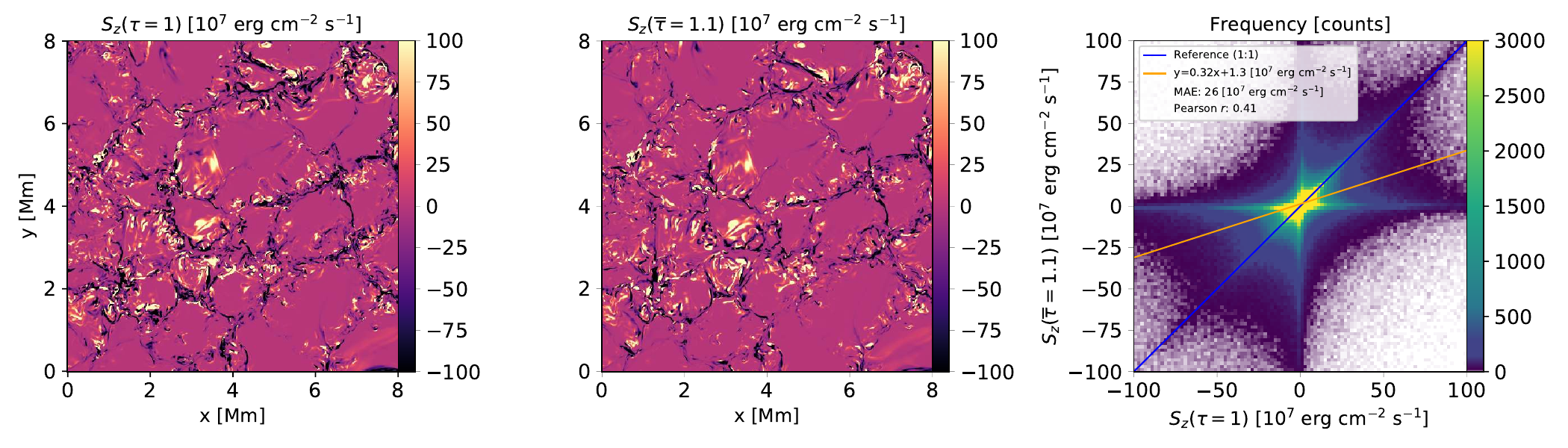}
		\caption{ \small  Vertical Poynting flux computed on optical surface $\tau=1$ (left panel) and geometrical surface $z=0$ Mm, where the spatially averaged optical depth is $\overline{\tau}=1.1$ (center panel). Notice the concentration of both positive and negative $S_z$ in IGLs. On the right panel is a 2-D histogram comparing pixel-by-pixel $S_z$ values computed on optical and geometrical surfaces. Note the different scales for the axes. }
		\label{fig:geom_opt}
    \end{center}
    
\end{figure}

\begin{figure}[ht]
    \begin{center}
        \includegraphics[width=1\linewidth]{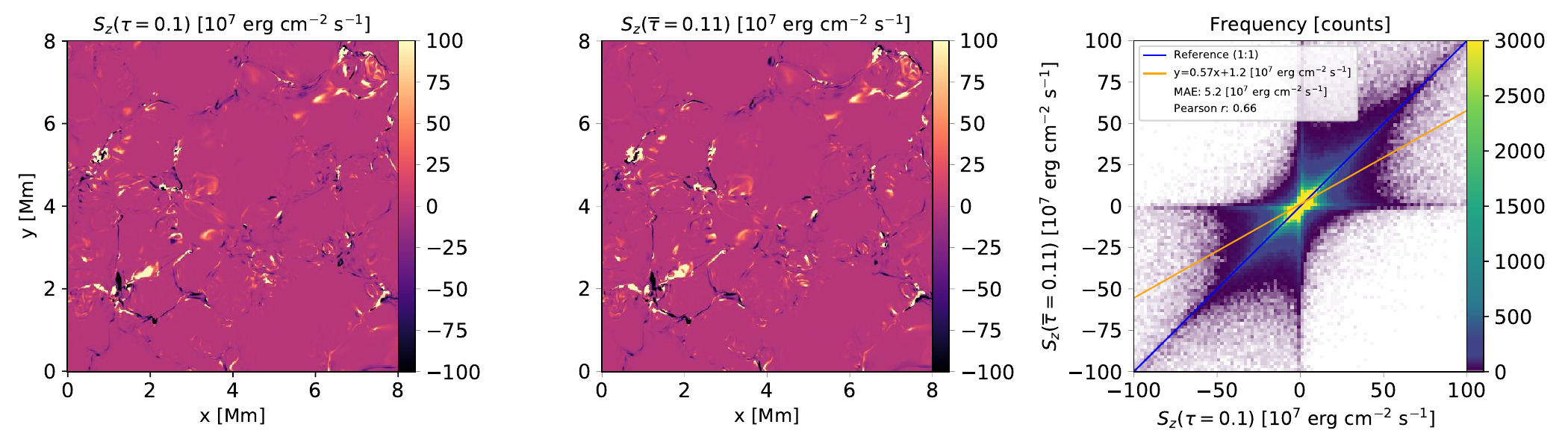}
		\caption{ Same as Figure \ref{fig:geom_opt}, but for the optical depth $\tau=0.1$ and geometrical surface $z=0.128$ Mm with spatially averaged optical depth $\overline{\tau}=0.11$. }
		\label{fig:geom_opt_01}
    \end{center}
    
\end{figure}

In their seminal paper, \citet{Withbroe1977} derived a threshold of upward energy flux from the photosphere that would be necessary to explain chromospheric and coronal heating in the quiet Sun -- $S_{z,thr}=4.3\times10^6$ erg cm$^{-2}$ s$^{-1}$.In MURaM simulations, vertical Poynting flux at $\overline{\tau}=1$ is just above the $S_{z,thr}$ value from \citep{Withbroe1977}. This is consistent with existing MURaM simulations, where hot corona is maintained by photospheric magnetoconvection \citep{Rempel2017,Breu2022,Breu2022Corrigendum}. However, we find that in IMaX observations, there is not enough Poynting flux whether we use the ideal-MHD method or PDFI\_SS, unless we consider only strong B-field pixels in the ideal-MHD case (Figure \ref{fig:Sz_estimates}). Furthermore, in the ideal-MHD case, vertical Poynting flux is lower than the minimum value required for heating by an order of magnitude. On the other hand, PDFI\_SS values are closer to $4.3\times10^6$ erg cm$^{-2}$ s$^{-1}$ in magnitude, but are frequently negative, indicating downward energy flux.

\begin{table*}
    \centering
    \begin{tabular}{l l c}
        Method & Target & $\overline{S_z}$ [erg cm$^{-2}$ s$^{-1}$] \\
        \hline
        \hline
        present work, ideal-MHD & observed QS, all pixels & $6.0\pm0.56\times10^5$  \\
        present work, ideal-MHD & observed QS, high B-field pixels & $1.1\pm0.087\times10^7$ \\
        present work, PDFI\_SS & observed QS & $-0.21\pm1.3\times10^6$ \\
        present work, MURaM & simulated QS, geometrical surface $z=0.000$ Mm ($\overline{\tau}=1.1$) & $4.36\times10^6$ \\
        present work, MURaM & simulated QS, geometrical surface $z=0.128$ Mm ($\overline{\tau}=0.11$) & $2.28\times10^7$ \\
        present work, MURaM & simulated QS, optical surface $\tau=1.0$ & $-3.11\times10^7$ \\
        present work, MURaM & simulated QS, optical surface $\tau=0.1$ & $1.94\times10^7$ \\
        \citet{Kazachenko2015}, PDFI\_SS & AR 11158 & $10^8-10^9$ \\
        \citet{Welsch2015}, ideal-MHD & AR 10930, plage & $\approx5\times10^7$ \\
        \hline
    \end{tabular}
    \caption{ Summary of photospheric Poynting flux estimates in the present work and in existing literature.}
    \label{tab:summary}
\end{table*}

What are the possible causes of discrepancies between different Poynting flux estimates, and why is Poynting flux negative in some of them? We propose several physical and methodological explanations. 

A non-trivial methodological issue with our analysis is the weak signal strength in the quiet Sun. This is particularly severe when it comes to Q and U Stokes vector signal strength, which adversely affects our inversions of ${\textbf{B}}_h$. Even in the ROI, where magnetic field strength exceeds 200 G -- a relatively high value for our data set -- there are significant discontinuities in the spatial distribution of horizontal magnetic field (right panel of Figure \ref{fig:dismabig}). These gaps affect both ideal-MHD and PDFI\_SS Poynting flux inversion methods. In ideal-MHD, as can be seen from Equation \ref{eqn:Sz}, Poynting flux is highly sensitive to ${\textbf{B}}_h$ as it appears in both terms of the expression. In PDFI\_SS, ${\textbf{B}}_h$ uncertainties affect both the ${\textbf{v}} \times {\textbf{B}}$ term as above and the spatial and temporal derivatives of the magnetic field.

Uncertainties in transverse magnetic field inversions $B_h$ propagate into issues with azimuth disambiguation. However, we see that with the IMaX signal strength, they do not meaningfully affect Poynting flux estimates, especially in the ideal-MHD scenario (Figure \ref{fig:randomSzs}). Instead, the emergence term $v_z B_h^2$ is responsible for virtually all signed Poynting flux and 99\% of unsigned Poynting flux. This is qualitatively consistent with some of the existing literature \citep[e.g., ][]{Liu2012}, but this fraction is much higher than in previous works. This is likely due to the weak magnetic field signal in our observational sample: For the shear term to be present, both linear and circular polarization signatures must be strong in the same pixel.

Transverse velocity inversions are another potential source of errors in Poynting flux inversions, though in the present work it is also likely to be a secondary order error. As discussed in \S\ref{sec:Vinv}, vorticity values inferred with DeepVel may be unreliable (Figure \ref{fig:vel_scatter}). Further, we do not have access to ``ground truth" when it comes to transverse velocity on the real Sun, and a neural network trained using supervised learning generates predictions that are only as good as the simulations they were trained on (from the relationship between continuum intensity and transverse flows, to the topology and magnitude of the flows). In MHD simulations, including STAGGER and MURaM, vortices are mostly concentrated in IGLs and have been shown to be spatially correlated with vertical Poynting flux in MURaM simulations \citep{Yadav2020,Yadav2021}. This, then, presents a clear avenue for improvement, particularly when DKIST observations with higher spatial resolution \citep[down to 0.03'', ][]{Rimmele2020} become available, since features in IGLs are especially vulnerable to resolution effects. Another way to improve the neural network approach is to train it to match coherence spectra, i.e. to match velocities at different frequencies in the Fourier space, as was done in \citet{Ishikawa2022}. 

Poynting flux inversions themselves can still be improved. We already explained how, unlike PDFI\_SS, the ideal-MHD method does not account for Poynting flux derived from $\displaystyle \frac{\partial \textbf{B}}{\partial t}$, but PDFI\_SS also has limitations. It has not been tested in the QS regime, particularly when only one polarity (negative in our case) of $B_z$ is present in the FOV.

Another explanation for negative vertical Poynting flux coulde be due to physical reasons. There are several pieces of evidence in favor of that possibility.

First, we are studying Poynting flux at the boundary layer between convection-dominated layers below the photosphere and radiation-dominated atmosphere. In such environment, it is reasonable to expect all energy fluxes (e.g. mass flux, convective flux) averaged over a representative FOV to become dominated by their horizontal components, which are mostly self-canceling, while vertical components approach near-zero values \citep{Steiner2008}. This is indeed the case in IMaX observations. Assuming ideal MHD conditions, we find that the horizontal Poynting flux ($|S_h|$) exceeds $|S_z|$ by a factor of $\approx 3$. \citet{Silva22} analyzed the same IMaX data set we use in our work and reported an even higher ratio of horizontal-to-vertical Poynting fluxes, likely due to higher velocity values in their inversions.

In MURaM simulations, we see that Poynting fluxes are principally concentrated in IGLs. An important corollary from this observation is that the emergence term of Poynting flux, which can only be negative provided $v_z<0$, is primarily negative in photospheric and chromospheric heights, which is indeed what we find (see Figure \ref{fig:muram_tauS}). Therefore, on average, the wave term of Poynting flux is larger in magnitude than the emergence term, in stark contrast with our findings in IMaX observations. As discussed above, this is likely due to a strong observational bias, wherein both IGL structure and magnetic concentrations that are not trivially oriented (neither completely parallel nor perpendicular to LOS) are subject to instrumental limitations. Observed ideal-MHD Poynting fluxes in IMaX, which are dominated by the emergence term yet positive, likely arise from magnetic concentrations located in granule interiors, such as those inside the ROI (see Figure \ref{fig:dismabig}). Analogs of such a structure can be seen in MURaM simulations as well, e.g. in Figures \ref{fig:muram_all} and \ref{fig:geom_opt} at $[x,y]=[3.5\text{ Mm},3.5\text{ Mm}]$ (just left and below of the center of FOV).

There are of course caveats when it comes to optical depth. First, it is unclear what optical depth corresponds to the formation of Fe I 5250.2 \AA\ line, which we used for spectropolarimetric inversions. It is evident that the line forms somewhere in the photosphere, but we are not aware of existing studies that looked at its response function. We find that this could be important, since in MURaM simulations average Poynting flux values are sensitive to changes in optical depth: from $\overline{\tau}=1.1$ to $\overline{\tau}=0.63$, $\overline{S_z}$ increases by a factor of 2.7 (see Figure \ref{fig:muram_tauS}). To see where the Fe I 5250.2 line could form, we calculate its response function using a MURaM atmospheric profile \citep{Rempel:2014:SSD}. We use one profile from granule and one from IGL, and in both cases (the latter is shown in Figure \ref{fig:muram_tauS}) we find that the line forms in the photosphere (300--400 km), but that it has a broad formation height range.

The second caveat related to optical depth is that a constant $\tau$ surface is very different from a constant height surface, and Poynting fluxes computed on optical surfaces deviate significantly from those computed on geometrical surfaces with comparable optical depth averaged over the FOV (see Figures \ref{fig:geom_opt} and \ref{fig:geom_opt_01}). In particular, while the average Poynting flux computed on the geometrical surface with $\overline{\tau}=1$ may be sufficient to match energy losses in the chromosphere and corona, the average Poynting flux computed on the geometrical surface $\tau=1$ is far from it, as it is $-3.11\times10^7$ erg cm$^{-2}$ s$^{-1}$ (Figure \ref{fig:muram_tauS}). Despite these differences, we treat all of our vector quantities' components as being either parallel or perpendicular to plane-of-sky (and line-of-sight). This can lead to unphysical results, since we are essentially dealing with vector projections and not true vectors. It is especially so in regions where optical surfaces are least aligned with geometrical ones, such as on the boundaries between granules and intergranular lanes. Incidentally, this is where 1) MURaM vertical Poynting flux is primarily concentrated, 2) transverse flows have the most shear and vorticity, and 3) resolution constraints have the highest detrimental effect \citep{Leka2022}.

\section{Conclusions} \label{sec:conc}

In this work we used two approaches -- the ideal-MHD and the PDFI\_SS methods -- to compute average photospheric Poynting fluxes from IMaX polarimetric observations. We tested several methods for deriving intermediate quantities required for computing Poynting flux. Principally, such quantities include the magnetic field azimuth, transverse velocities, and electric fields. We also looked at the outputs of 3-D radiative MHD code MURaM between $\tau=10^9$ and $\tau=-5\times10^{-8}$ to glean insights from simulated photospheric data.

Our quantitative estimates of Poynting flux do not reveal a consistent picture with respect to whether photospheric Poynting flux is sufficient to explain chromospheric and coronal heating (Table \ref{tab:summary}). However, we can outline several important findings:
\begin{enumerate}
    \item The ideal-MHD approach yields ambiguous estimates of $S_z$, but this could be explained by the quality of available data. When considering only pixels with relatively high magnetic field values ($|B|>50$ G), the resulting average Poynting flux suffices to explain chromospheric heating, but $S_z$ averaged over all pixels does not (Figure \ref{fig:Sz_estimates}). It is possible that, due to instrumental limitations, we miss on many of the small and/or transient magnetic concentrations;
    
    \item The $180^{\circ}$ azimuthal ambiguity barely affects the estimates of the ideal-MHD approach (Figure \ref{fig:randomSzs}). This is because Poynting fluxes derived via the ideal-MHD method are dominated by the emergence term $v_z B_h^2$. This could also explain the lack of Poynting flux when averaged over all pixels. The importance of the emergence term has been reported before, but it is likely exaggerated in our results, since pixels with both vertical and transverse magnetic field (which are both necessary to produce the wave term $(\textbf{v}_h \cdot \textbf{B}_h) B_z$ of Poynting flux) are difficult to detect in QS magnetograms. Indeed, in MURaM simulations, the wave term is on average positive and larger in magnitude than the emergence term, which is concentrated in IGLs and, consequently, is negative on average. When more advanced observations are available, such that the bias against the wave term is diminished and resolving azimuthal ambiguity becomes relevant, we point to our Poynting flux optimization method as a way to disambiguate azimuths while meaningfully constraining $S_z$;

    \item Poynting flux obtained with the PDFI\_SS method is highly time-dependent, insufficient for chromospheric and coronal heating, and is negative in many of the frames in our time series (Figure \ref{fig:Sz_estimates}). It is also sensitive to azimuth disambiguation. The variability and sensitivity to magnetic field azimuthal orientation can be caused by the reliance of PDFI\_SS on spatial and temporal derivatives, combined with the noisy data sample. The closeness of it to zero can be attributed to the photosphere being a boundary layer between convection-dominated subsurface and radiation-dominated lower atmosphere;

    \item MURaM simulations also display vertical Poynting flux that flips sign around $\tau=1$ and is dominated by (unsigned) horizontal Poynting flux, supporting the boundary layer explanation (Figure \ref{fig:muram_tauS}). $S_z$ in MURaM simulations is frequently negative around the $\tau=1$ surface, particularly in IGLs (see Figures \ref{fig:muram_all} and \ref{fig:geom_opt}). At the same time, the upward Poynting flux is more than sufficient to explain chromospheric and coronal heating. While it may look like Poynting flux is close to the heating threshold around $\tau=1$ (Figure \ref{fig:muram_tauS}), this region is in the deep photosphere, where the sign of Poynting flux flips and below the formation height of most observable spectral lines. It should be noted that MURaM simulations that extend into the corona \citep{Rempel2017} produce a self-maintained QS corona (about 1.5 Million K) with sufficient Poynting flux. However, those simulations have lower resolution and the Poynting flux comes more from braiding of QS network field that is mostly absent in our simulation set. MURaM simulations of coronal loops also show that photospheric energy output is sufficient to maintain hot corona \citep{Breu2022,Breu2022Corrigendum}.
    
\end{enumerate}

The main question -- whether observed QS photosphere produces enough magnetic energy in the form of Poynting flux to heat the chromosphere and corona -- remains open. There are, however, promising signs that this uncertainty will be cleared up in the future. DKIST observations, particularly with VBI, DL-NIRSP, and VTF instruments, can be used to observe photospheric magnetic fields with unprecedented polarization sensitivity, resolution, and cadence \citep{Rimmele2020}. Repeating this analysis with DKIST data is one of the most obvious avenues for future work.

We can also improve our methodology moving forward, particularly as it pertains to transverse magnetic field inversions, including azimuth disambiguation, and transverse velocity inversions. For the former, a physics-based approach such as ME0 is preferable to the more stochastic or optimizing approaches used in this work. We can also use acute angle disambiguation, provided we have QS observations sufficiently far from the disk center. We may also achiever higher fidelity in transverse magnetic field inversions by using an inversion scheme that solves for magnetic filling factor \citep{Leka2022}. For transverse velocity inversions, modifying DeepVel so that it is trained to match vorticity as well as velocity can be useful, since Poynting flux is associated with shear flows and vortices in IGLs.

Finally, numerical MHD simulations present a convenient avenue of exploring relationships between time- and height-dependent upward flux of magnetic energy and different structures in the quiet Sun. This area has remained largely unexplored, due to a lack of observational counterparts with which to verify potential findings, but it is more relevant now, in the era of DKIST. For an investigation of Poynting flux that is more directly comparable to observations, observables such as Stokes vectors must be computed using forward models and then inverted. It should be noted that it is unclear whether such an approach will result in physical values, since inversions produce quantities on optical surfaces where vector cross products are not meaningful. However, an approach involving forward modeling can be used to assess the model fidelity and, by extension, whether it can be used to make useful Poynting flux predictions. Detailed and focused studies of numerical simulations are therefore necessary.

\section*{Acknowledgements} 
This work is supported by NASA FINESST award 20-HELIO20-0004. This material is based upon work supported by the National Center for Atmospheric Research, which is a major facility sponsored by the National Science Foundation under Cooperative Agreement No. 1852977. The authors thank Anna Malanushenko for general comments on the paper and K.D. Leka for her valuable advice on azimuth disambiguation. The authors also thank the anonymous referee for useful suggestions, particularly in regards to azimuth disambiguation.

\facilities{SUNRISE (IMaX).}

\software{\href{https://github.com/jaimedelacruz/pyMilne}{pyMilne} \citep{delaCruzRodriguez2019},
          \href{https://www.cora.nwra.com/AMBIG/ME0.html}{ME0} \citep{Leka2009},
          \href{https://pyflct.readthedocs.io/en/latest/index.html}{FLCT} \citep{Fisher2008},
          \href{https://github.com/aasensio/deepvel}{DeepVel} \citep{AsensioRamos2017},
          \href{http://cgem.ssl.berkeley.edu/cgi-bin/cgem/PDFI_SS/index}{PDFI\_SS} \citep{Fisher2020},
          MURaM \citep{Rempel:2014:SSD}.}

\bibliography{main}{}
\bibliographystyle{aasjournal}

\end{document}